\documentstyle[preprint,prb,aps]{revtex}

\begin{document}
\draft

\title{Electron Depletion Due to Bias of a T-Shaped Field-Effect Transistor}
\author{G. A. Georgakis and Qian Niu}
\address{Department of Physics, University of Texas at Austin,
Austin, Texas 78712}
\date{\today}
\maketitle

\begin{abstract}
A T-shaped field-effect transistor, made out of a pair of
two-dimensional electron gases, is modeled and
studied.  A simple numerical model is
developed to study the electron distribution
vs. applied gate voltage for different gate lengths.
The model is then improved to account for depletion and
the width of the two-dimensional electron gases.
The results are then compared to the experimental ones and
to some approximate analytical calculations and are found to be in good
agreement with them.
\end{abstract}



\narrowtext
\section{Introduction}
\label{intro}
Extensive work has been published on various nano-structure devices that use
two-dimensional electron gases (2DEG)~\cite{kirk,ando}.  Numerous devices
have been
built using nanofabrication that involve shaping the 2DEG by
applying gate voltages.
Some of the techniques use in-plane
gating~\cite{wieck,pothier}; however,  most of them use
off-plane gating~\cite{kastner,ishii,stormer}.
Here, we study an off-plane device whose gate is also a 2DEG.

Our work was prompted by a paper by H. L. Stormer et al.~\cite{stormer}  In
that paper, a new type of T-shaped field-effect transistor~(T-FET)
was presented with an ultra short
gate length.  The transistor is made out of a pair of 2DEGs that are oriented
perpendicular to each other.  One is
used as the source-drain channel, and the other is used as the gate. When
the gate voltage exceeds a certain critical value, a depletion
region is created
in the source-drain channel which cuts off the current flow through it.

The device is made exclusively using molecular beam epitaxy (MBE).  The
2DEG of
the gate is made by standard modulation doping of a $GaAs$ substrate in the
following way:\\
\begin{tabular}{llll}
$4.3\ \mu m$ &	undoped	& $Al_{0.3}Ga_{0.7}As,$\\
$200$ \AA &	undoped	& $GaAs$\ quantum well,\\
$150$ \AA &	undoped	& $Al_{0.3}Ga_{0.7}As$ spacer,\\
$1.5\times 10^{16}\ m^{-2}$ & $Si\ \delta$ doping, & \\
$20\ \mu m$ &	undoped	& $Al_{0.3}Ga_{0.7}As.$
\end{tabular}\\
Following this, the structure is now cleaved on one edge, and the growth
continues in a perpendicular direction.  The growth sequence is as follows:\\
\begin{tabular}{llll}
$200$ \AA &	undoped	& $Al_{0.3}Ga_{0.7}As$\ separator,\\
$150$ \AA &	undoped	& $GaAs$\ quantum well,\\
$150$ \AA &	undoped	& $Al_{0.3}Ga_{0.7}As$\ spacer,\\
$2\times 10^{16}\ m^{-2}$ & $Si\ \delta$ doping, & \\
$1700$ \AA &	undoped	& $Al_{0.3}Ga_{0.7}As,$\\
$600$ \AA &	undoped	& $GaAs.$ \\
\end{tabular}\\
This produces a modulation-doped quantum well which is $200$~\AA\ wide
for the
gate and $150$~\AA\ wide for the source-drain channel.  A cross section of
the transistor is shown in Fig.~\ref{cross}.

In Sec.~\ref{theory}, we develop a very simple model for the
transistor, based on work done by X. Liu and one of us~\cite{xiangming}.  In
this model, we assume that the electron distribution is related to the
potential
energy by the Thomas-Fermi equation
\equation
\label{thomas}
n = (\delta-U)\nu.
\endequation
Here, $n$ is the extra electron distribution due to the applied gate voltage,
$\nu = \frac{m^*}{\pi \hbar^2}$ is the two-dimensional density of states, and
$\delta$ is the change in chemical potential of the two 2DEGs due to the
bias.
The equation is valid everywhere except inside the
insulating barriers and any created
depletion regions.  It does not give an
exact solution to the problem, but it can
give a good understanding of the general behavior of this kind of
semiconductor
device. Furthermore, we initially assume that the two 2DEGs have no width in
the direction in which they are confined by the quantum wells, and that the
donors that created the gases are in the same position as the 2DEGs.  We also
assume that no depletion region is created.  These assumptions allow
us to come up with an analytic solution to the problem using Fourier
transforms.

In Sec.~\ref{depletion}, we allow for a depletion
region to be created.  The solution using Fourier transforms is no longer
feasible, and a numerical solution is used throughout.  Special
techniques had
to be developed for this.

In Sec.~\ref{width}, we make a more realistic model of the transistor
by taking into account the finite width of the gases.
The techniques used to solve
these are very similar to those used in Sec.~\ref{depletion}.

In Sec.~\ref{analytic}, an approximate analytic approach is used for the
problem inspired by Ref.~\onlinecite{glazman}.  The results of the two
methods are then compared in the conclusion.

\section{The onset of depletion}
\label{theory}
As a first approach, we assume that the two 2DEGs have no width and that no
depletion region is created.  Furthermore, we assume that the
donors are in the same position as the 2DEGs.  As shown in Fig.~\ref{xy},
we choose the $x$ axis for the gate and the $y$ axis for the source-drain
channel.
The above assumptions simplify the equations
for the potential energy of the gases to
\begin{eqnarray}
\lefteqn{U(x,y) = } \nonumber \\
& - & \frac{e^2}{2\pi \epsilon} \int_a^\infty dx'\, n_x (x')\, \log
\left|(x-x')^2 + y^2\right|^{1/2} \nonumber \\
& - & \frac{e^2}{2\pi \epsilon} \int_{-\infty}^\infty dy'\, n_y
(y')\, \log \left|x^2 + (y-y')^2\right|^{1/2},
\end{eqnarray}
which on each axis gives
\begin{mathletters}
\label{uxy}
\begin{eqnarray}
\label{ux}
\lefteqn{U(x)\equiv U(x,0) = } \nonumber \\
& - & \frac{e^2}{2\pi \epsilon} \int_a^\infty dx'\, n_x (x')\, \log
\left|x-x'\right| \nonumber \\
& - & \frac{e^2}{2\pi \epsilon} \int_{-\infty}^\infty dy'\, n_y
(y')\, \log \left|x^2 + y'^2\right|^{1/2},
\end{eqnarray}
and
\begin{eqnarray}
\label{uy}
\lefteqn{U(y)\equiv U(0,y) = } \nonumber \\
& - & \frac{e^2}{2\pi \epsilon} \int_a^\infty dx'\, n_x (x')\, \log
\left|x'^2 + y^2\right|^{1/2} \nonumber \\
& - & \frac{e^2}{2\pi \epsilon} \int_{-\infty}^\infty dy'\, n_y
(y')\, \log \left|y-y'\right|.
\end{eqnarray}
\end{mathletters}
In addition, for each of the 2DEGs the Thomas-Fermi equations give
\begin{mathletters}
\equation
n(x) = [\delta_x-U(x)]\nu,
\endequation
and
\equation
n(y) = [\delta_y-U(y)]\nu
\endequation
\end{mathletters}
for the $x$ and $y$ axis respectively.
Substituting in for the potential energy terms on each axis from
Eqs.~(\ref{uxy}), and rescaling a few of the parameters we get
\begin{mathletters}
\label{nxy}
\begin{eqnarray}
\label{nx}
n_x(x) = \delta_x & + & \frac{2}{\pi} \int_a^\infty dx'\, n_x
(x')\, \log \left|x-x'\right| \nonumber \\
& + & \frac{2}{\pi} \int_{-\infty}^\infty dy'\, n_y (y')\, \log
\left|x^2 + y'^2\right|^{1/2},
\end{eqnarray}
and
\begin{eqnarray}
\label{ny}
n_y(y) = \delta_y & + & \frac{2}{\pi} \int_a^\infty dx'\, n_x
(x')\, \log \left|x'^2 + y^2\right|^{1/2} \nonumber \\
& + & \frac{2}{\pi} \int_{-\infty}^\infty dy'\, n_y (y')\, \log
\left|y-y'\right|.
\end{eqnarray}
\end{mathletters}
The new units used are:
length:~$a_0 = \frac{4\pi \epsilon \hbar^2}{m^* e^2}$;
density:~$\frac{m^*eV_g}{\pi\hbar^2}=\frac{eV_g}{E_F}n_0$;
voltage:~$eV_g$.
These are two integral equations with two unknown functions $n_x(x)$ and
$n_y(y)$.  In addition to the above equations, the two additional
requirements
are the neutrality
\equation
\label{neutr}
\int_a^\infty dx\, n_x(x) + \int_{-\infty}^\infty dy\, n_y(y) = 0,
\endequation
and the fact that
\equation
\label{dxy}
\delta_x - \delta_y = 1.
\endequation
The above equation states the fact that the difference in the
chemical potentials of the two sides is just the gate voltage.

We can obtain one equation for $n_x(x)$ by using Fourier transforms to
eliminate $n_y(y)$.  The resulting integral equation for $n_x(x)$ is
\begin{eqnarray}
\label{nxf}
\lefteqn{n_x(x) = 1 -\frac{2}{\pi} \int_a^\infty dx'\, n_x(x')\, }
\nonumber\\
& & \times \left\{ \log \left| \frac{x+x'}{x-x'}\right| +
\int_0^\infty dk\, \frac{e^{-(x+x')k}}{k+2} \right\}.
\end{eqnarray}
In a very similar way, we can obtain an equation for $n_y(y)$ in terms
of $n_x(x)$.  By manipulating Eqs.~(\ref{nxy}), we can express
the Fourier transform of $n_y(y)$ in terms of $n_x(x)$ as
$\hat{n}_y(k_y) = 2\pi\delta_y\frac{|k_y|\delta(k_y)}{|k_y|+2}
 -2\int_a^\infty dx'\, n_x(x')\,
\frac{e^{-|k_y|x'}}{|k_y| + 2} $
which, after taking the inverse Fourier transform, gives
\equation
\label{nyf}
n_y(y) = - \frac{2}{\pi} \int_a^\infty dx'\, n_x(x')\,
\int_0^\infty dk_y \frac{\cos(k_yy)e^{-k_yx'}}{k_y+2}.
\endequation

Before trying to solve this equation, we investigate the long range solution.
The numerator of the $dk$ integral in Eq.~(\ref{nxf}) approaches zero
quickly as $x\rightarrow \infty$. Assuming
that $n_x(x) \sim \frac{\alpha}{x^\beta}$ as $x\rightarrow \infty$, and
by setting $t = x/x'$, we must have
$1=\frac{2}{\pi} \frac{\alpha}{x^{\beta-1}} \int_{a/x}^\infty
dt'\, \frac{1}{t^\beta}\, \log \left| \frac{1+t}{1-t}\right|.$
Since $\int_{0}^\infty dt'\, \frac{1}{t}\, \log \left|
\frac{1+t}{1-t}\right| = \frac{\pi^2}{2}$,
we can infer that $\alpha=\frac{1}{\pi}$ and $\beta=1$. This gives
\equation
n_x(x) = \frac{1}{\pi x}, \ \ x\rightarrow \infty
\endequation
for the asymptotic form.

For the asymptotic form of $n_y(y)$, we need to manipulate
Eq.~(\ref{nyf}).  We first set
$k'=k|y|$ which, along with the condition that $y \rightarrow \infty$,
enables us to compute the $dk$ integral so that we now have
$n_y(y)=-\frac{2}{\pi}\int_{a}^{\infty}dx'\frac{1}{x'^2+y^2}$.  The
$dx'$ integral can now be computed by using the variable substitution
$t=x'/y$.  The final result for the asymptotic form of $n_y(y)$ is
\equation
n_y(y) = \frac{1}{2\pi |y|}, \ \ |y|\rightarrow \infty
\endequation

Eq.~(\ref{nxf}) can be solved either by successive iterations or by using
a matrix method, since it is linear.  The result is shown in
Fig.~\ref{nxyxy}a.
$n_y(y)$ can be computed from Eq.~(\ref{nyf}) since $n_x(x)$ is now
known numerically.  The result is shown in Fig.~\ref{nxyxy}b.

If we put all the right units back into the result for the
$n_y(y)$ distribution,
we can estimate what gate voltage can start the formation of the
depletion region
on the $y$ axis.  This happens when the distribution at $y=0$ becomes
equal to
the background density $n_{0y}$.  The condition is
\equation
\label{vg}
\frac{e V_{g_{c}}}{E_F} = -\frac{1}{n_y(y=0)}.
\endequation
In this case, $V_{g_{c}}$ comes out to be $230\ mV$.

This result is not in  very good agreement with the experimental
result~\cite{stormer}.
We give two reasons to explain this discrepancy.  First, our
critical voltage value represents the onset of depletion,
which is not exactly when the source-drain channel is cut
off. Electrons can still tunnel through if the depletion width is on the
order of the Fermi wavelength.  Another, less important reason,
is the behavior of the
2DEG density close to the boundary.  The wave function for
the 2DEG can be represented by
$\psi (x, z) = \frac{1}{\sqrt{L_z}}e^{k_zz}\sqrt{\frac{2}{L_x}}\sin k_xx,$
where $L_x$ and $L_z$ are chosen boundaries for the gas.  Under this
assumption, we can calculate the density of the gas to be
\begin{equation}
\frac{n_x(x)}{n_{0x}} = 1-\frac{{\rm J}_1(2k_Fx)}{k_Fx}.
\end{equation}
The effect of this form of
charge distribution is to give an effective  gate length, which is
slightly larger than the geometrical one.  To get an estimate for the
difference, we calculate the mean of the extra charge distribution which
turns out to be
\[ \int_0^\infty du\, u\, \frac{{\rm J}_1(2u)}{u}=\frac{1}{2},\ u=k_Fx.\]
This is an addition of about $30$~\AA\ or $0.31\ a_0$ to the
geometrical gate length.
We return to this again in Sec.~\ref{width}.

\section{Dependence of the depletion on the bias voltage}
\label{depletion}

The Fourier method used in the previous section can not be used when a
depletion region is present.  The Thomas-Fermi equations do not hold in that
case.   A numerical solution had to be employed by discretizing $n_x(x)$ and
$n_y(y)$, converting the integral equations into matrix equations, and
simultaneously solving for the charge distribution everywhere.

Similar to the treatment employed in Sec.~\ref{theory}, we first
try to investigate the behavior of
the distribution for large $x$ and $y$.  We assume that $n_x(x) \sim
\frac{\alpha}{x}$ and that $n_y(y) \sim \frac{\beta}{x}$ in this limit.
First, by considering charge neutrality from Eq.~(\ref{neutr}), we must have
$\int_a^l dx\, \frac{\alpha}{x} + \int_{-l}^l dy\, \frac{\beta}{|y|}=0$,
which implies (as $l \rightarrow \infty$) that $\beta = - \frac{\alpha}{2}$.
Similarly to Sec.~\ref{theory}, we obtain from the limit
$\delta_x = \frac{3 \pi \alpha}{4}$ and $\delta_y = - \frac{\pi \alpha}{4}$,
which combined with the requirement of Eq.~(\ref{dxy}) gives
\begin{equation}
\label{dxylr}
\delta_x = \frac{3}{4},\ \
\delta_y = - \frac{1}{4}
\end{equation}
and
\begin{equation}
n_x(x) = \frac{1}{\pi x},\ x \rightarrow \infty ;\
n_y(y) = - \frac{1}{2 \pi |y|},\ |y| \rightarrow \infty.
\end{equation}
The same results were obtained in Sec.~\ref{theory}.

To accommodate depletion, only small changes need to
be made to Eqs.~(\ref{nxy}) and Eq.~(\ref{neutr}).
Assuming a depletion region from $-d/2$ to $d/2$, the new equations
are
\begin{mathletters}
\label{nxyd}
\begin{eqnarray}
\label{nxd}
n_x(x) = \delta_x & + & \frac{2}{\pi} \int_a^\infty dx'\, n_x
(x')\, \log \left|x-x'\right| \nonumber \\
& + & \frac{4}{\pi} \int_{0}^{d/2} dy'\, n_y (y=d/2)\, \log
\left|x^2 + y'^2\right|^{1/2}\nonumber \\
& + & \frac{4}{\pi} \int_{d/2}^{\infty} dy'\, n_y (y')\, \log
\left|x^2 + y'^2\right|^{1/2}
\end{eqnarray}
and
\begin{eqnarray}
\label{nyd}
n_y(y) = \delta_y & + & \frac{2}{\pi} \int_a^\infty dx'\, n_x
(x')\, \log \left|x'^2 + y^2\right|^{1/2} \nonumber \\
& + & \frac{4}{\pi} \int_{0}^{d/2} dy'\, n_y (y=d/2)\, \log
\left|y^2-y'^2\right|^{1/2}\nonumber \\
& + & \frac{4}{\pi} \int_{d/2}^{\infty} dy'\, n_y (y')\, \log
\left|y^2-y'^2\right|^{1/2}.
\end{eqnarray}
\end{mathletters}

Eqs.~(\ref{nxyd}) proved to be a real challenge to solve numerically.  The
natural logarithms go to $\pm \infty$ when either $x$ or $y$ approaches $\pm
\infty$ or $0$.  Infinities are also encountered whenever $x=x'$ or $y=y'$.
To solve these equations the following technique was employed:  A cut-off $l$
was introduced on all the integrals.  The integrals now have two main parts.
One from $0$ to $l$ (or $-l$ to $l$) and another from $l$ to $\infty$ (or
this
and $-\infty$ to $-l$). It turns out that we were able to obtain an
analytical
result to the second portion of these
 integrals assuming that for large enough $l$, the $n_x(x)$
and $n_y(y)$ distributions approach their long range approximations.
The latter
parts of these integrals become tails that we attach to the integrals.
The attached tails are
\begin{mathletters}
\label{txy}
\begin{eqnarray}
\label{tx}
t_x(x) & = & \frac{2}{\pi} \int_l^\infty dx'\, \left( \frac{1}{\pi
x'} \right)\, \log \left| x-x' \right| \nonumber \\
& + & \frac{4}{\pi} \int_{l}^\infty dy'\, \left(
\frac{-1}{2\pi |y'|} \right)\, \log \left|x^2 + y'^2\right|^{1/2}
\nonumber \\
& = & \frac{2}{\pi^2} \int_0^{x/l} dt\, \frac{1}{t} \log
\left|\frac{(1-t)^2}{1+t^2} \right|^{1/2}
\end{eqnarray}
with the latter result achieved by setting $t = x/x'$ in the $dx'$ integral,
and $t = x/y'$ in the $dy'$ integral.  Similarly, we obtain
\begin{eqnarray}
\label{ty}
t_y(y) & = & \frac{2}{\pi} \int_l^\infty dx'\, \left( \frac{1}{\pi
x'} \right)\, \log \left| x'^2 + y^2 \right|^{1/2} \nonumber \\
& + & \frac{4}{\pi} \int_{l}^\infty dy'\, \left( \frac{-1}{2\pi
|y'|} \right)\, \log \left|y^2-y'^2\right|^{1/2} \nonumber \\
& = & \frac{2}{\pi^2} \int_0^{y/l} dt\, \frac{1}{t} \log
\left|\frac{1+t^2}{1-t^2} \right|^{1/2}.
\end{eqnarray}
\end{mathletters}
With the addition of these tails and the $l$ cut-offs,
Eqs.~(\ref{nxyd}) become
\begin{mathletters}
\label{nxydt}
\begin{eqnarray}
\label{nxdt}
n_x(x) = \delta_x & + & \frac{2}{\pi} \int_a^l dx'\, n_x
(x')\, \log \left|x-x'\right| \nonumber \\
& + & \frac{4}{\pi} \int_{0}^{d/2} dy'\, n_y (y=d/2)\, \log
\left|x^2 + y'^2\right|^{1/2}\nonumber \\
& + & \frac{4}{\pi} \int_{d/2}^{l} dy'\, n_y (y')\, \log
\left|x^2 + y'^2\right|^{1/2}\nonumber \\
& + & \frac{2}{\pi^2} \int_0^{x/l} dt\, \frac{1}{t} \log
\left|\frac{(1-t)^2}{1+t^2} \right|^{1/2}
\end{eqnarray}
and
\begin{eqnarray}
\label{nydt}
n_y(y) = \delta_y & + & \frac{2}{\pi} \int_a^l dx'\, n_x
(x')\, \log \left|x'^2 + y^2\right|^{1/2} \nonumber \\
& + & \frac{4}{\pi} \int_{0}^{d/2} dy'\, n_y (y=d/2)\, \log
\left|y^2-y'^2\right|^{1/2}\nonumber \\
& + & \frac{4}{\pi} \int_{d/2}^{l} dy'\, n_y (y')\, \log
\left|y^2-y'^2\right|^{1/2}\nonumber \\
& + & \frac{2}{\pi^2} \int_0^{y/l} dt\, \frac{1}{t} \log
\left|\frac{1+t^2}{1-t^2} \right|^{1/2}.
\end{eqnarray}
\end{mathletters}
The neutrality equation (Eq.~(\ref{neutr})) also changes to
\begin{eqnarray}
\label{neutrdt}
\int_a^\l dx\, n_x(x)
&+& 2\int_{0}^{d/2} dy\, n_y(y=d/2) \nonumber\\
&+& 2\int_{d/2}^{l} dy\, n_y(y) = 0.
\end{eqnarray}

Eqs.~(\ref{nxydt}) can now be solved with the additional requirement of
Eq.~(\ref{dxy}).
A matrix solution is obtained by discretizing
$n_x(x)$ and $n_y(y)$ and solving for those values together
with $\delta_x$ and $\delta_y$.
The solutions for $n_x(x)$ and $n_y(y)$ are shown in Fig.~\ref{nxyxy}.
The corresponding
solutions for Eq.~(\ref{dxy}) are $\delta_x = 0.749$ and $\delta_y = 0.251$,
which are in very good agreement with the predictions of Eqs.~(\ref{dxylr}).

\section{Adding width to the electron gases}
\label{width}

To get a more realistic result from our simple model, we now try to add
width to the 2DEGs.  Assuming that
the 2DEGs are in the ground state and extending from $-g/2$
to $g/2$, the wave function for either one is then given by
$ \psi(x) = \sqrt{\frac{2}{g}} \cos (\frac{\pi x}{g}) $.
To make the numerical calculations easier, we assume a square
distribution and
pick the width $f$ so that
$\int_{-f/2}^{f/2} dx\, x^2\, \frac{1}{b} =
\int_{-g/2}^{g/2} dx\, x^2\, \frac{2}{g} \cos^2 (\frac{\pi x}{g})$,
which gives $f = \sqrt{1-\frac{6}{\pi^2}}\ g$.
The two quantum wells for the 2DEGs are $200$~\AA\ on the
$x$ axis and $150$~\AA\
on the $y$ axis, which give $b = 1.268\, a_0$ and $c = 0.951\, a_0$ for the
$x$ and $y$ axis square width respectively.
The new setup is shown in Fig.~\ref{xy}.

With the addition of width, Eqs.~(\ref{nxydt}) become:
\widetext
\begin{mathletters}
\label{nxydtw}
\begin{eqnarray}
\label{nxdtw}
n_x(x) = \delta_x
& + & \frac{2}{\pi} \int_a^l dx'\, n_x (x')\, \frac{1}{b}
\int_{-b/2}^{b/2} dy' \log \left|(x-x')^2 + y'^2\right|^{1/2}\nonumber \\
& + & \frac{4}{\pi} \int_{0}^{d/2} dy'\, n_y (y=d/2)
\frac{1}{c} \int_{-c/2}^{c/2} dx'
\log \left|(x-x')^2 + y'^2\right|^{1/2}\nonumber \\
& + & \frac{4}{\pi} \int_{d/2}^l dy'\, n_y (y')
\frac{1}{c} \int_{-c/2}^{c/2} dx'
\log \left|(x-x')^2 + y'^2\right|^{1/2}\nonumber \\
& + & \frac{2}{\pi^2} \int_0^{x/l} dt\, \frac{1}{t} \log
\left|\frac{(1-t)^2}{1+t^2} \right|^{1/2}
\end{eqnarray}
and
\begin{eqnarray}
\label{nydtw}
n_y(y) = \delta_y
& + & \frac{2}{\pi} \int_a^l dx'\, n_x (x')\, \frac{1}{b}
\int_{-b/2}^{b/2} dy' \log \left|x'^2 + (y-y')^2\right|^{1/2}\nonumber \\
& + & \frac{4}{\pi} \int_{0}^{d/2} dy'\, n_y (y=d/2)
\frac{1}{c} \int_{-c/2}^{c/2} dx'
\log \left|x'^2+x'(y^2+y'^2)^{1/2}+(y^2-y'^2)\right|^{1/2}\nonumber \\
& + & \frac{4}{\pi} \int_{d/2}^l dy'\, n_y (y')
\frac{1}{c} \int_{-c/2}^{c/2} dx'
\log \left|x'^2+x'(y^2+y'^2)^{1/2}+(y^2-y'^2)\right|^{1/2}\nonumber \\
& + & \frac{2}{\pi^2} \int_0^{y/l} dt\, \frac{1}{t} \log
\left|\frac{1+t^2}{1-t^2} \right|^{1/2}.
\end{eqnarray}
\end{mathletters}
\narrowtext
The above equations can be solved in a manner similar to that used in solving
Eqs.~(\ref{nxydt}).  The results are shown in Fig.~\ref{nxyxy}.
Fig.~\ref{depl} shows a plot of the depletion width vs. applied voltage
for different values of the gate length.  Fig.~\ref{vga} shows a plot
of the applied voltage needed to create a given depletion width vs. the
gate length for various depletion widths.

We now try to again compare these results with the experimental ones.
The experimental critical voltage is $450\ mV$
which corresponds to $\frac{eV_c}{E_{Fy}}\approx 50$, shown on
Fig.~\ref{vga}.
Assuming an effective gate length of about $3\ a_0$, this corresponds to
a depletion length of approximately $4.5\ \lambda_{Fy}$.
Fig.~\ref{volt} shows the voltage distribution on the $y$ axis for various
depletion widths.

\section{An approximate analytic approach}
\label{analytic}

Following work done by D.~B.~Chklovskii~et~al.~\cite{glazman},
we make the assumption that the potential is constant across the 2DEGs,
just like it is in metals.
{}From now on, we will refer to this as the metal
approximation.  The
depletion region can be modeled by a continuous positive charge
distribution.  This is shown in Fig.~\ref{pot},
along with all the different boundary conditions.
To obtain the potential everywhere, we use conformal mapping to transform
to a
region where potential calculations will be easier.  In this section, we
switch the labels for the $x$ and $y$ axis to make the conformal mapping
transformation easier.
The transformation to do this is
\equation
w = \sin^{-1} (z/l);\ z = x + iy,\ w = u + iv.
\endequation
Under this transformation, $-l$ and $l$ go to $-\pi/2$ and $\pi/2$
respectively,
and $a$ goes to $a' = \sinh^{-1}(a/l)$.  The boundary condition along the
depletion region now becomes
\equation
\label{bc}
- \left. \frac{\partial \Phi}{\partial v} \right|_{v=0^+} +
  \left. \frac{\partial \Phi}{\partial v} \right|_{v=0^-}  =
\frac{e n_0}{\epsilon} l\cos u.
\endequation

We examine the simpler case, where $a=0$, which implies that
 $a'$ in the transformed
coordinates is now also zero.  From symmetry, we need only to consider the
region $0 \leq u \leq \pi/2$.  The potential is
\begin{equation}
\Phi_+(u,v) = -V_g + \frac{2V_g}{\pi}u + \sum_{m=\rm even}A_m \sin(mu)\,
e^{-mv}
\end{equation}
and
\begin{equation}
\Phi_-(u,v) = \sum_{k=\rm odd}B_k \cos(ku)\, e^{kv}
\end{equation}
above and below the $u$ axis respectively.

By applying the boundary condition of Eq.~(\ref{bc}) and the fact that
the two potentials have to match at the boundary, we can get equations
for the $A$ and $B$ coefficients.  The equation for the $B$
coefficients is
\begin{eqnarray}
\sum_{k=\rm odd}\frac{1}{m-k}\left( \frac{eB_k}{E_F}\right) &=&
\frac{2}{m^2-1}\left( \frac{d}{a_0}\right) \nonumber\\
&-& \frac{1}{m}\left( \frac{eV_g}{E_F}\right);\  {m=\rm even}.
\end{eqnarray}
It turns out that the electric field, in the $x$ direction along the
boundary, has a singularity when $x=l$ (Ref.~\onlinecite{glazman}).  This can
be canceled by the right choice of $l$, which is what determines
the relation between the applied voltage and the depletion length.  In
the $uv$ coordinates, the electric field $E_x$, along the $x$
axis, transforms to
\begin{equation}
-\left.\frac{\partial\Phi}{\partial x}\right|_{y=0, x\rightarrow l}=
-\lim_{u\rightarrow \pi/2}\frac{1}{l\cos u}
\left.\frac{\partial\Phi}{\partial u}\right|_{v=0, u\rightarrow\pi/2}.
\end{equation}
To cancel the $1/l\cos u=1/\sqrt{l^2-x^2}$ singularity, we must have
$\left. -\frac{\partial \Phi}{\partial v}\right|_{v=0, u\rightarrow\pi/2}
= 0$,
which gives the condition
\begin{equation}
\sum_{k=\rm odd} (-1)^{(k+1)/2} k \left( \frac{eB_k}{E_F}\right) = 0.
\end{equation}
A numerical solution is employed to solve for the depletion region width
vs. applied voltage.  The result is
\begin{equation}
\frac{d}{a_0} = \frac{2}{\pi}\frac{eV_g}{E_F}.
\end{equation}
The above equation predicts a linear relationship between the applied voltage
and depletion length.  It also predicts that $V_{g_{c}}=0$. This is
shown in Fig.~\ref{depl}.  The $2/\pi$ slope is in exact agreement with the
numerical results.

Finally, we try to investigate another configuration by using the metal
approximation.

We assume that the transistor has another gate on
the other side, so that the new configuration is now in the shape of a
``cross''.  This adds more symmetry to the problem, which actually makes
the solution under the metal approximation easier.
We now only have to consider the upper half
of the $uv$ plane.  The new solution for $a=a'=0$ is
\begin{equation}
\frac{d}{a_0} = \frac{eV_g}{E_F}.
\end{equation}
This gives a $\pi/2$ improvement on the slope of the graph of depletion
versus voltage over the ``T'' configuration.
We also calculate $d$ vs.
$V_g$ for $a\ne 0$.  These results are shown in Fig.~\ref{crossfig}.
The results are in very good agreement with
the high bias limit and predict the correct slope between the
applied voltage and the depletion width.
It should be noted that all the slopes converge to the same value
in the limit of high bias voltage.  Of course, the ``cross'' transistor
is not a device that can be manufactured with current technology.
However, the study of this interesting configuration showed how much
improvement
an additional mirror gate can give.

\section{Conclusion}
\label{conclusion}

The Thomas-Fermi equation allows us to get a good sense of the general
behavior of depletion formation on the ``T'' transistor.
Because of its simplicity, the metal approximation used in
Sec.~\ref{analytic}
provided a very powerful method for checking our results.  Such
approximation can
also be used to get a general idea of a problem without having  to
explore complicated analytical or numerical solutions.
And though the method does not
predict the correct minimum voltage needed to start the formation of
the depletion region, it does give the correct slope and
shows good agreement for large bias voltage.

\section*{Acknowledgments}
We would like to thank Xiangming Liu for his help and guidance
with some of the numerical problems in this paper and
L. I. Glazman for his useful discussions on the metal approximation.
Also for their useful discussions, G. A. G. would like to thank
M. C. Chang, Ertugrul Demircan, and R. Jahnke.
This work was supported by the Welch Foundation.


\begin{figure}
\caption
{Schematic cross section of the T-FET.  The two 2DEGs are shown in
the gray shaded areas.}
\label{cross}
\end{figure}

\begin{figure}
\caption
{The system of coordinates chosen for the model.  We always assume
that the positive donors are found where the 2DEGs (shaded areas) are.}
\label{xy}
\end{figure}

\begin{figure}
\caption
{The electron distributions for the gate ($x$ axis) and
source-drain ($y$ axis) channels are shown for different depletion
lengths $d$
and with or without assumed width for the 2DEGs.
Lengths are in units of $a_0$, and
electron distributions are in units of $\frac{eV_g}{E_F}n_0$.}
\label{nxyxy}
\end{figure}

\begin{figure}
\caption{Plots of the depletion length $d$ as a function of
applied voltage $V_g$ for different gate lengths $a$.  From left to
right, the solid lines correspond to gate lengths $\frac{a}{a_0}=0,1,...,8$.
The dotted line corresponds to the prediction of the metal approximation
for $\frac{a}{a_0}=0$.}
\label{depl}
\end{figure}

\begin{figure}
\caption{A plot of the applied voltage $V_g$, required to produce a given
depletion width $d$, as a function of the gate length $a$.  From bottom
to top, the solid lines correspond to depletion widths
$\frac{d}{a_0}=0,5,10,15,20\ (5\ a_0=1\ \lambda_{Fy})$.  The dotted line
corresponds to
the experimental critical voltage of $450\ mV$.}
\label{vga}
\end{figure}

\begin{figure}
\caption{A plot of the potential energy $U_y(y)$, in units of the Fermi
energy, along the $y$ axis for a depletion
$\frac{d}{a_0}=5,10,15,20\ (5\ a_0 = 1\ \lambda_{Fy})$.
The dotted line represents the Fermi level.}
\label{volt}
\end{figure}

\begin{figure}
\caption{The boundary conditions for the potential in both the $xy$ and $uv$
coordinate systems.  The transformation is $w = \sin^{-1} (z/l)$, with
$z=x+iy$ and $w=u+iv$.  The solid lines are the positive background charge,
and the dotted lines are the 2DEGs, which are assumed to be at a constant
potential.}
\label{pot}
\end{figure}

\begin{figure}
\caption{Depletion length $d$ vs. applied voltage $V_g$ for the ``cross''
transistor using
the metal approximation for different gate lengths $a$.  From left to
right the solid lines correspond to $\frac{a}{a_0}=0,1,...,5$.}
\label{crossfig}
\end{figure}


\begin{references}

\bibitem{kirk}
{\it Nanostructures Physics and Fabrication}, edited by M. A. Reed and
W. P. Kirk (Academic, New York, 1989)

\bibitem{ando}
T. Ando, A. B. Fowler, and F. Stern
{\it\rmp} {\bf 54,} 437 (1982).

\bibitem{wieck}
A. D. Wieck and K. Ploog,
{\it\apl} {\bf 56,} 928 (1990).

\bibitem{pothier}
H. Pothier, J. Weis, R. J. Haug, K. v. Klitzing, and K. Ploog,
{\it\apl} {\bf 62,} 3174 (1993).

\bibitem{kastner}
M. A. Kastner,
{\it\rmp} {\bf 64,} 849 (1992).

\bibitem{ishii}
Masami Ishii, Kazuhiko Matsamuto, Hidehiro Morozumi, Yoshinobu Sugiyama, and
Tsuneroni Sakamoto,
{\it Jpn. J. Apll. Phys.} {\bf 32,} {\bf L}36 (1993).

\bibitem{stormer}
H. L. Stormer, K. W. Baldwin, L. N. Pfeiffer, and K. W. West,
{\it\apl} {\bf 59,} 9 (1991).

\bibitem{xiangming}
X. Liu, and Q. Niu,
{\it\prb} {\bf 46,} 16 (1992).

\bibitem{glazman}
D. B. Chklovskii, B. I. Shklovskii and L. I. Glazman
{\it\prb} {\bf 46,} 7 (1992).

\end{references}
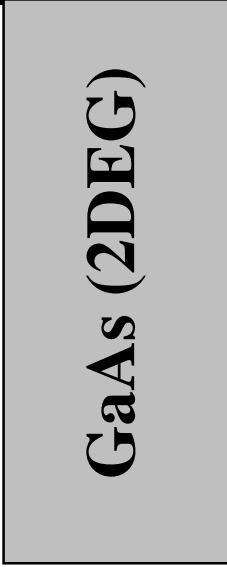
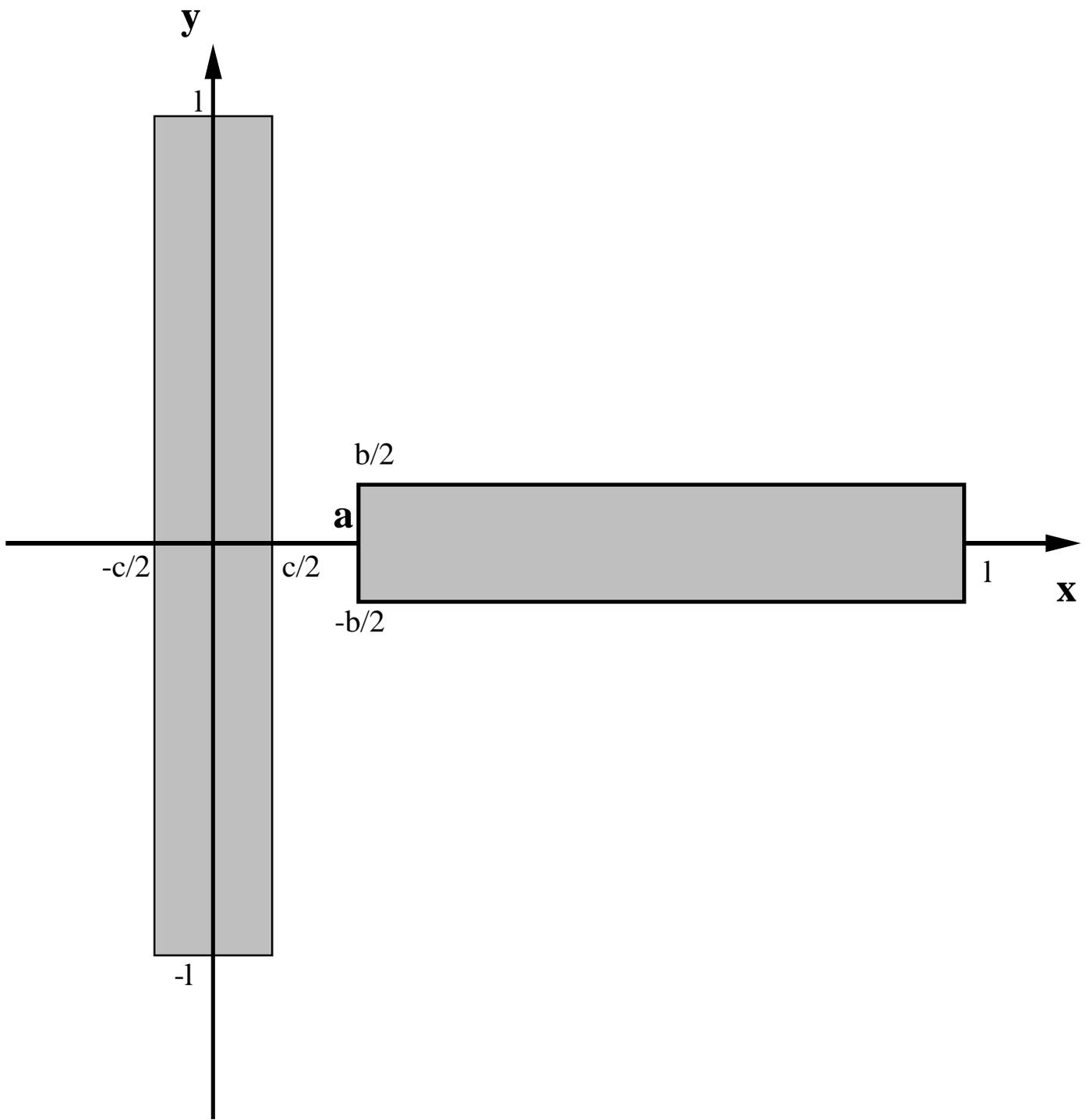
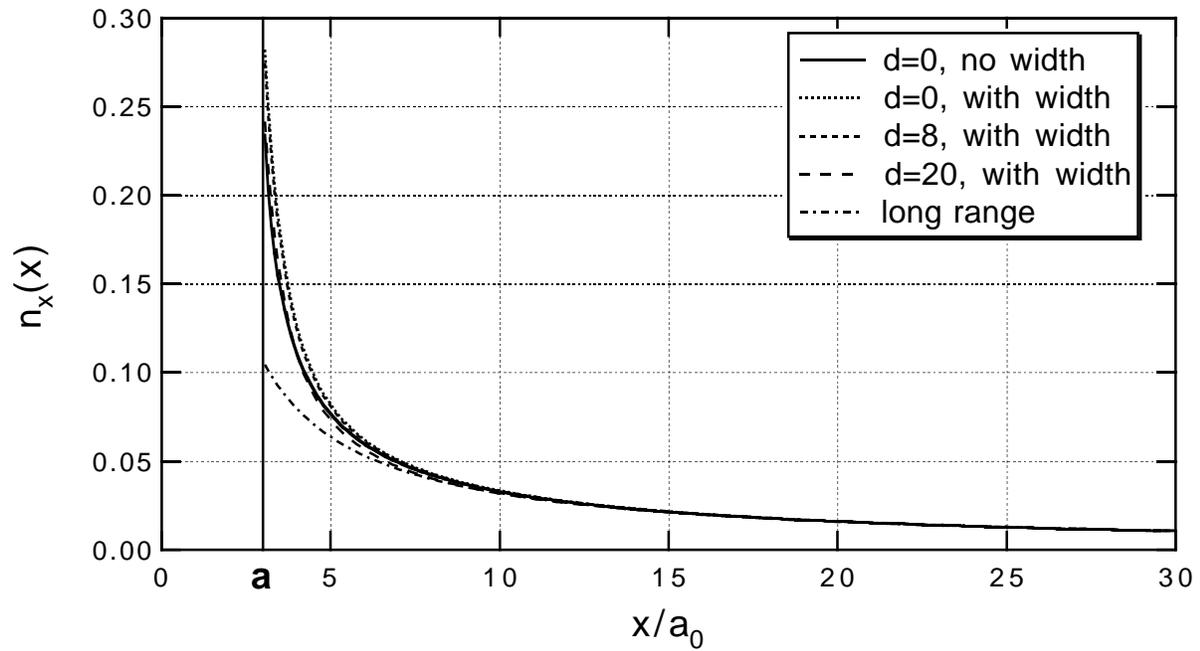
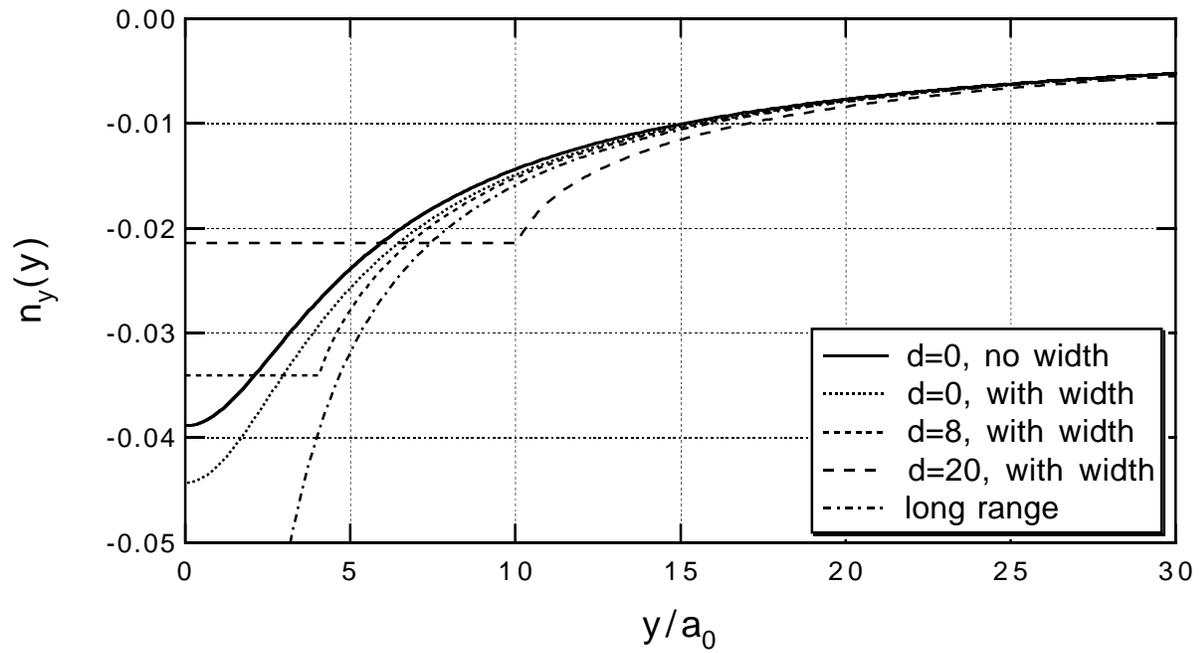
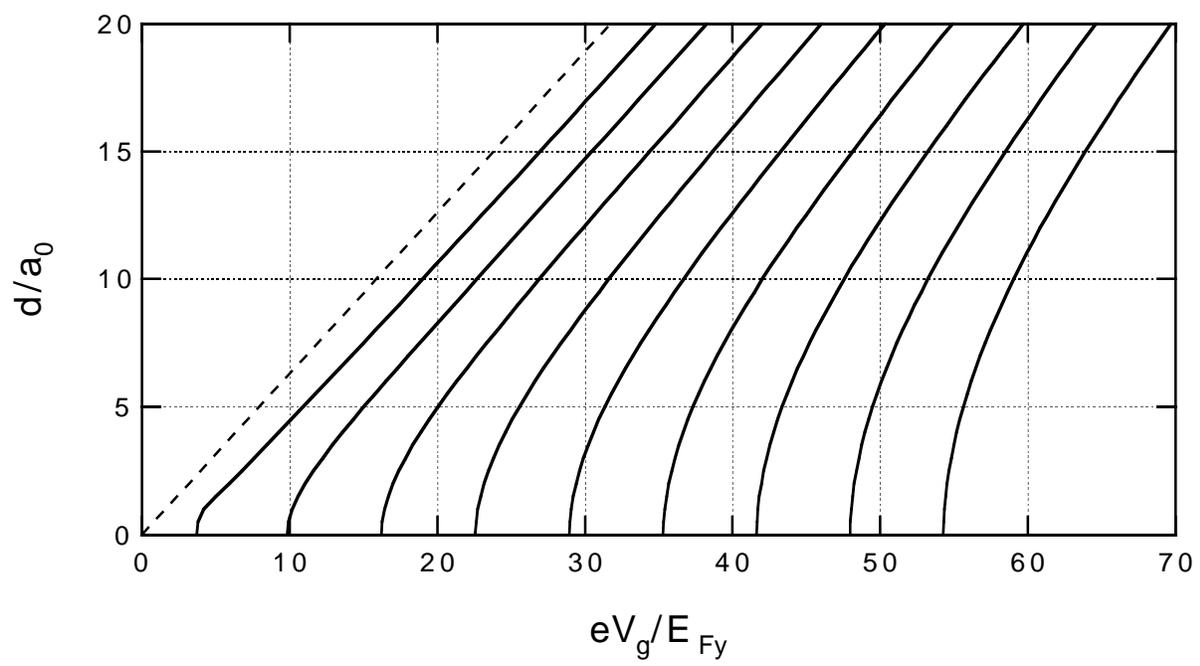
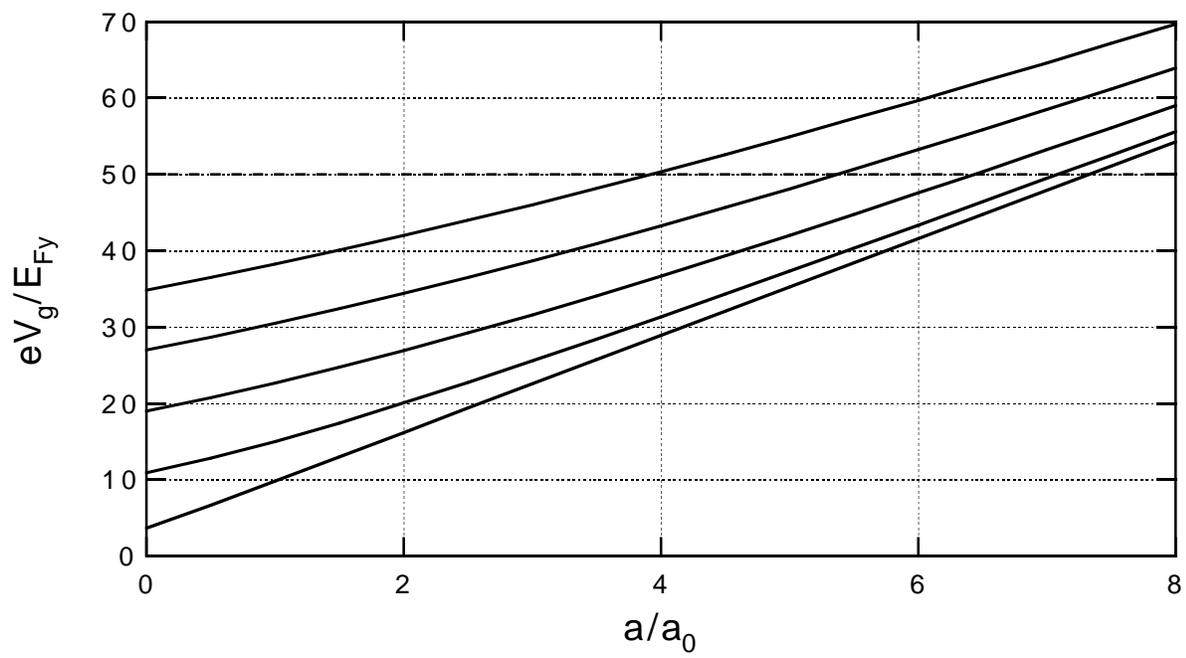
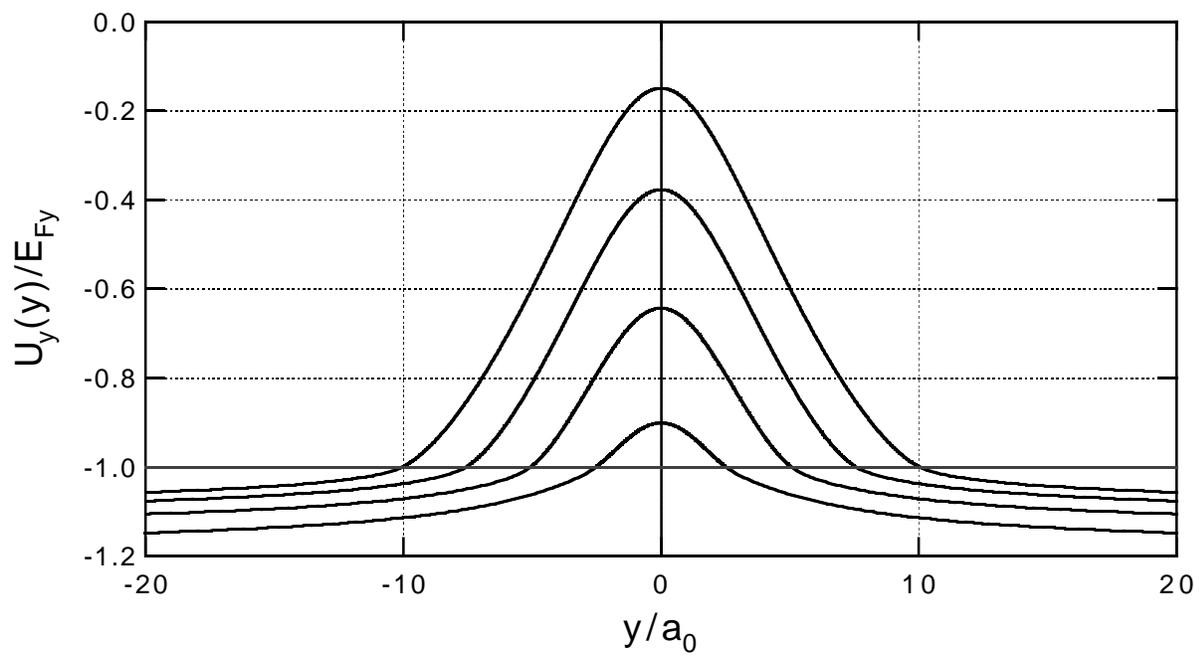
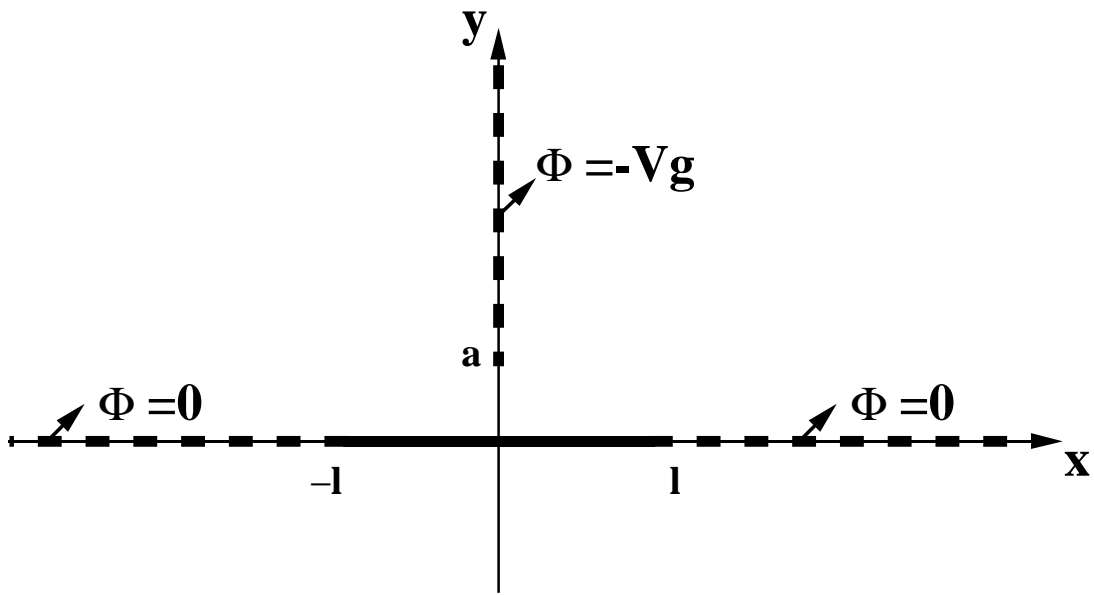
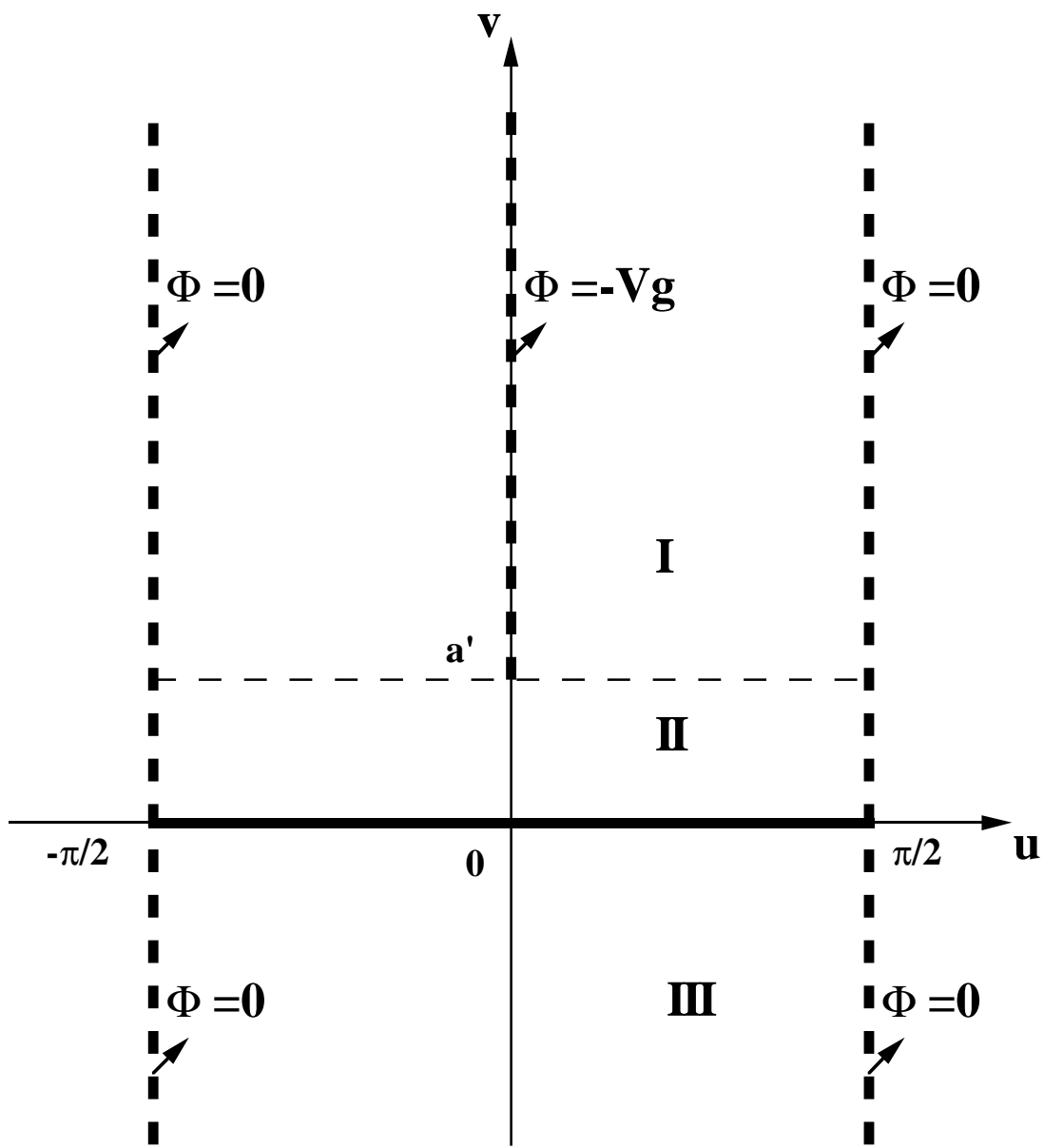
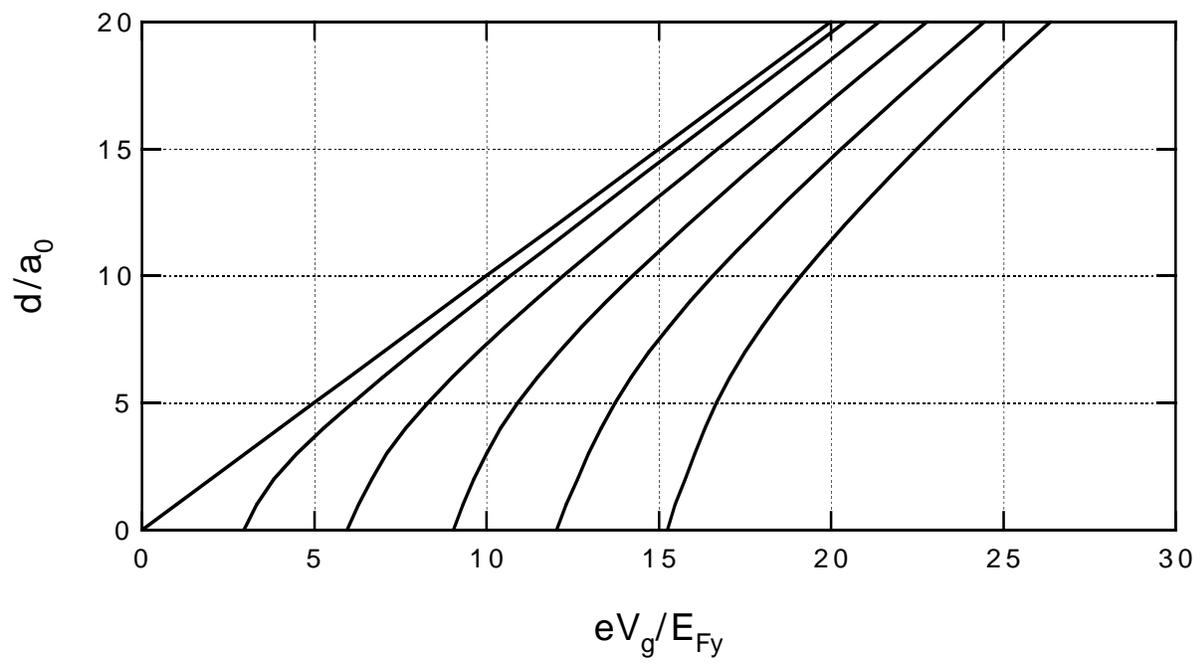
\end{document}